# 3D Analytics:

# Opportunities and Guidelines for Information Systems Research

Gunther Gust[1,*], Tobias Brandt[2], Otto Koppius[3], Markus Rosenfelder[4], Dirk Neumann[4]

1 Center for Artificial Intelligence and Data Science (CAIDAS), University of Würzburg, gunther.gust@uni-wuerzburg.de

2 European Research Center for Information Systems (ERCIS), University of Münster

3 Rotterdam School of Management (RSM), Erasmus University Rotterdam

4 Information Systems Research, University of Freiburg

Abstract. Progress in sensor technologies has made three-dimensional (3D) representations of the physical world available at a large scale. Leveraging such 3D representations with analytics has the potential to advance Information Systems (IS) research in several areas. However, this novel data type has rarely been incorporated. To address this shortcoming, this article first presents two showcases of 3D analytics applications together with general modeling guidelines for 3D analytics, in order to support IS researchers in implementing research designs with 3D components. Second, the article presents several promising opportunities for 3D analytics to advance behavioral and design-oriented IS research in several contextual areas—such as healthcare IS, human-computer interaction, mobile commerce, energy informatics and others. Third, we investigate the nature of the benefits resulting from the application of 3D analytics, resulting in a list of common tasks of research projects that 3D analytics can support, regardless of the contextual application area. Based on the given showcases, modeling guidelines, research opportunities and task-related benefits, we encourage IS researchers to start their journey into this largely unexplored third spatial dimension.

## 1    Introduction

During the past decade, progress in information and sensor technologies has enhanced our ability to translate the physical world into data. Examples range from airborne drones that provide high-resolution three-dimensional (3D) digital representations of entire swaths of land (Manfreda et al. 2018), through digital twins of our cities and neighborhoods that support critical urban services (Fuller et al. 2020, Fan et al. 2021), to wearable devices that digitize human movement to capture health-related information (Bajracharya et al. 2019, Bardhan et al. 2020). The commonality between these applications is the translation of both our three-dimensional physical environment and human behavior within that environment into the virtual space.

As a discipline, Information Systems (IS) research is well-acquainted with this concept as it relates to two-dimensional representations. For instance, geographical information systems (GIS) have long been integral components of the IS portfolio of many companies (Mennecke et al. 2000, Farkas et al. 2016), particularly in the energy and logistics sectors (Gust et al. 2016, Sarkar 2007), while map-based interfaces are the centerpieces of many services that dominate the digital economy, such as Airbnb and shared-mobility services (Brandt and Dlugosch 2021). On top of this, the Covid-19 global pandemic has emphasized the importance of GIS in public health management and decision-making as well as in communicating current developments and measures. Similar to other recent advances in data collection and analysis technologies, such as those related to neuroscience (Dimoka et al. 2011), the increasing availability and ubiquity of 3D data adds a layer of complexity to this context, often requiring advanced methodologies to tackle, but also a layer of opportunity for IS research.

In this article, we outline the opportunities and challenges associated with 3D analytics—which we understand as the combination of 3D data with a toolbox of appropriate methods and processing techniques to derive insights and value from it. We continue in Section 2 by defining 3D analytics, presenting previous work and providing background information on sources of 3D data and 3D data formats. In Section 3, we focus on the technologies and practices of 3D analytical modeling: We first present two brief showcases that illustrate the value of 3D analytics and can serve as blueprints for IS researchers designing 3D analytics applications. Thereafter, we use the showcases to derive general modeling guidelines that support researchers in successfully implementing research designs with 3D components. In Section 4, we then focus on the opportunities of applying 3D analytics in IS research. For this purpose, we first review the value potentials of applying 3D analytics in IS design, analytics and decision support, as well as behavioral IS research in several contextual areas, such as human-computer interaction, healthcare IS, mobile commerce and others. Second, we look at common tasks of IS researchers across contextual areas (such as data collection and experiment design) and explain how 3D analytics can support in executing such tasks. We conclude with a brief summary.

## 2 3D Analytics

### 2.1 Definition

We define for the purpose of this paper 3D analytics as the *practices and technologies* for transforming *3D data into value*. A dataset qualifies as *3D data* if it contains three-dimensional geometric coordinates $(x, y, z)$, i.e. coordinates that refer to locations in Euclidean space[1]. Although multiple perspectives on analytics exist (Holsapple et al. 2014), we perceive analytics as the *practices and technologies* for the transformation of data into value. This transformation covers at least one of the steps of the information value chain, i.e. the sequence of turning data into information, knowledge, decisions, actions, and eventually value (Abbasi et al. 2016, Sharma et al. 2014). The practices and technologies facilitate analytical modeling, which typically follows the steps of data collection, preparation, feature engineering, modeling, and evaluation—as described in KDD, CRISP-DM and other, related analytical modeling frameworks (Fayyad et al. 1996, Shearer 2000).

### 2.2 Formats and Sources of 3D Data

3D data is generally represented in one of two fundamental data formats, 3D point clouds or 3D polygon meshes. These are characterized in Table 1. Even though a large number of 3D file formats exist, these are often application specific and are built upon these two basic formats. Point clouds are the raw data format generated by most technologies that capture 3D data from real-world environments. Among those, most prevalent technologies include light-detection-and-ranging (LiDAR) systems in planes and autonomous vehicles, but also accelerometers in wearables and handheld devices. 3D polygon meshes are a more complex data format, consisting of 3D points, connecting edges and surfaces. Except for artificially-created environments (such as virtual reality), the polygon meshes are often a product of data transformations, e.g. from 3D point clouds or from multiple photographic images via photogrammetry.

**Table 1 3D data formats and associated data sources**

| Data Format | Illustration | Description | Data sources |
|---|---|---|---|
| 3D point cloud | 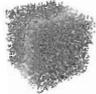 | Points with coordinate triples $(x, y, z)$ | Light detection and ranging (LiDAR) |
| | | Optionally with color information (red, green, blue) | Time-of-flight (TOF) cameras |
| | | | Accelerometers (in wearables and handhelds) |

---

[1]*Euclidean* describes the fundamental space of classical geometry, as first described by Euclid (Eves (1963)).

| | | | |
|---|---|---|---|
| 3D polygon meshes | 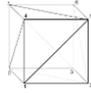 | Points, edges, surfaces (can be represented in different ways) | Virtual environments<br>Stereo-photogrammetry<br>Often also derived from 3D point clouds (e.g. 3D city models) |

The aptitude of the 3D data formats and 3D data capturing technologies, including their advantages and disadvantages, depend on the task that 3D analytics should address and the domain of the research project. We return to these advantages and disadvantages when discussing the opportunities of 3D analytics in Section 4.

## 2.3 Prior Work

As sources of 3D data are becoming ubiquitous, techniques of analyzing 3D data have already been exploited by several disciplines. Accordingly, this paper is not the first review on the opportunities offered by 3D analytics. 3D analytics has become particularly prominent in the geoscience and forestry disciplines (Dong and Chen 2017), where 3D models of large geographic regions are generated, e.g., for inferring terrains and vegetation. Similarly, urban planning research is taking advantage of 3D city models for a variety of applications (Biljecki et al. 2015), such as mitigating the effects of noise and air pollution resulting from urban traffic. On a more granular scale, 3D building models greatly facilitate planning and operations in the construction sector (Wang and Kim 2019).

Apart from these application areas, prior reviews and research agendas have also focused on specific steps of 3D data processing. For instance, computer vision research has for a long time been focusing on the task of generating 3D data, e.g., on human behavior and the physical environment, out of images for subsequent usage (Cyganek and Siebert 2011, Shirai 2012, Aggarwal and Xia 2014). Machine learning researchers are currently exploring methods for learning from 3D data formats (Bronstein et al. 2017, Griffiths and Boehm 2019). Finally, visualization research has been evaluating the effectiveness of 3D representations (e.g., in 3D plots) for conveying information more effectively to human users in a variety of tasks (Çöltekin et al. 2016).

In summary, technologies to capture and process 3D data are on the rise in many disciplines; however, there is little prior research on the corresponding opportunities for IS research. This article aims to close this gap by describing the potentials of 3D analytics in several contextual areas of IS research, including both behavioral and design-oriented research traditions. Before returning to these opportunities, we next focus on the practices of 3D analytical modeling—in order to support IS researchers in conducting projects involving 3D components.

## 3   3D Analytical Modeling

The aim of this section is to provide guidelines for conducting research projects using 3D analytics. For this purpose, we first create a typology of 3D analytics problems in Section 3.1. We then provide two 3D analytics showcase studies in Sections 3.2 and 3.3, which demonstrate the benefits of 3D analytics and can serve as blueprints for IS researchers aiming to design 3D analytics applications. We then use the typology to review the showcases and derive modeling guidelines in Section 3.4.

### 3.1   Typology of 3D Analytics Problems

The typology in Figure 1a assists in structuring 3D analytics problems and allows us later to develop differentiated modeling guidelines. It disentangles the modeling of real-world phenomena into the modeling of entities and relations. Entities describe objects—such as buildings, a TV screen, or the human body. Relations specify how the entities are related to each other. In 3D analytics, relations typically depend on distance, but may also go beyond distance measures, e.g. whether an entity is in line-of-sight or in field-of-vision of another entity. The typology also distinguishes 3D analytics from 2D and 1D spatial analytics as well as non-spatial analytics: While the physical representations of entities and relations are always 3D in the real world, the given application examples show that researchers may model entities and relations in lower dimensionalities. To fall under the definition of 3D analytics, however, at least one—either entity or relation—has to be modeled in three dimensions.

For the application examples in Figure 1a, the nature of entities and relations as well as their modeling dimensionalities are given in more detail in Figure 1b. In both our showcases, we later use 3D entities and 3D relations; however, this is not always necessary. Bramah et al. (2018), for example, study the injury risk of runners, which they relate to the vertical movement of the hips during running (colloquially known among runners as 'hip drop'). For this analysis, the authors model the runners' hips as 3D entities, from which they derive the vertical height difference between the hips as a 1D relation, which then corresponds to their variable of interest, namely the hip drop. A converse example is a study on location-based advertisement by Ghose et al. (2019), which analyzes the impact of location-based advertising in a large, multistory shopping mall by targeting ads according to users' trajectory patterns through the mall. The authors thereby model shopper trajectories as 3D relations and reduce the shoppers themselves to 1D entities. For some applications, it may also be reasonable to model different types of entities in different dimensionalities. For example, in a human-computer interaction experiment, Leroy et al. (2013) study users' attention preferences in second screen applications. Thereby, the authors model screens as 2D entities and the viewer's head as a 3D entity. In summary, several combinations of modeling dimensionalities for entities and relations exist, even for similar applications. Deciding on adequate modeling dimensionalities is one of the challenges of applying 3D analytics, which we later address in our modeling guidelines.

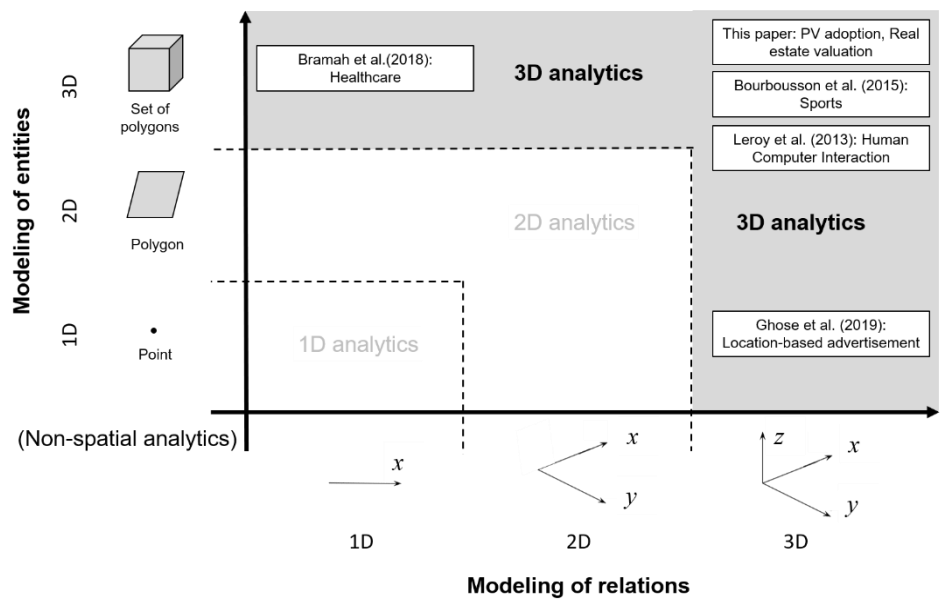

a) Typology (with exemplary applications)

| Application | Entities | Relations |
|---|---|---|
| Real-estate valuation (this paper) | Building volume (3D) | Location (3D) |
| PV adoption (this paper) | Roof orientation, inclination, size (3D) | PV potential (3D) |
| Healthcare (Bramah et al. 2018) | Hips (3D) | Hip drop (1D) |
| Sports (Bourbousson et al. 2015) | Basketball players (3D) | Field of vision (3D) |
| Location-based advertisement: (Ghose et al. 2019) | Customers (1D) | Trajectories through a multi-level shopping mall (3D) |
| Human-computer interaction (Leroy et al. 2013) | User's head (3D) and TV screen (2D) | Line-of-sight of viewer (3D) |

b) Modeling dimensionalities of entities and relations of exemplary applications

**Figure 1 Typology of 3D analytics problems and modeling dimensionalities**

## 3.2 Showcase: Valuation of Real Estate

Understanding social dynamics in urban environments is an issue of great interest to IS researchers in the area of smart cities (Brandt et al. 2018, Rosemann et al. 2021). While there are many elements to those dynamics, one important indicator is the value of real estate. According to hedonic pricing theory, the value of real estate comprises characteristics of the building and of the surrounding environment. Following this theory, we expect 3D building models to be useful for the valuation of real estate, since 3D building geometries contain information about several building characteristics—such as building height, size, and layout. However, 3D geometries have scarcely been used for real estate valuation. The few previous efforts that utilize 3D geometries (Helbich et al. 2013, Yu et al. 2007) rely to some extent on manual data collection, limiting the theoretical insights regarding the analytical modeling procedures and their practical applicability for large-scale real estate valuation.

Our study mimics the business requirements of real estate investors who want to estimate potential revenues from renting apartments based on historical rent prices in their business area. To do so, we predict future rent prices of apartments based on observed prices of the past within the same area. We locate the study in the city of Berlin in order to obtain a sufficiently large sample of historical rent prices. We structure the study along the common steps of analytical modeling: study design and data collection, feature engineering, method and model selection, and model evaluation (Shearer 2000, Fayyad et al. 1996, Shmueli and Koppius 2011).

*Study design and data collection*. Since 3D geometries should describe buildings more precisely than two-dimensional footprints or one-dimensional point data, we decide to model the building entities in 3D. The decision about the modeling of the relations is more ambiguous. The common practice in real estate valuation is to utilize 2D relations for describing the proximity to amenities (such as shopping centers, transportation access, and schools). Some approaches, however, use 3D relations, for instance to model the view from apartments (Yu et al. 2007) or the incoming solar irradiance (Helbich et al. 2013). While capturing these 3D relations requires considerable modeling effort (such as incorporating the position of windows to determine the view), the magnitude of their effect on prices is less clear. We therefore strike a balance between modeling effort and accuracy and use simple 3D relations that include only the latitudinal and longitudinal coordinates of the building along with its elevation above sea level. This design follows Martins-Filho and Bin (2005), who found that the elevation can serve as a proxy for views and is much simpler to model.

*Feature engineering.* Regarding the relations, we follow the previous argumentation and derive the geo-coordinates of the apartment (encoded in the features *latitude*, *longitude*, and *elevation*) to describe the 3D characteristics of the environment. In addition, we utilize the 2D relation *urban district*, referring to the one out of 12 urban districts of Berlin in which the apartment is located. Regarding the entities, we create the high-level feature *building volume* to summarize the 3D geometries of the building. This feature describes the size of the building and also provides information about the number of additional apartments within the building. The features *apartment size* and *number of rooms* describe the 2D and 1D geometries of the entities. The outcome variable, *apartment rent*, is non-spatial. In sum, this yields the variables given in Table 2.

**Table 2 Features and types in real estate showcase[2]**

| Feature | Type | Value range | Explanation |
| --- | --- | --- | --- |
| *elevation* [m] | 3D relation | Continuous | Elevation (above sea level) of the apartment building |
| *latitude* | 3D relation | Continuous | Geocoordinates |
| *longitude* | 3D relation | Continuous | |
| *urban district* | 2D relation[a] | {1,…,12} | District of the apartment building |
| *building volume* [m$^3$] | 3D entity | Continuous | Volume of the apartment building |
| *apartment size* [m$^2$] | 2D entity | Continuous | Size of the apartment footprint |
| *number of rooms* | 1D entity | Continuous | Number of rooms of the apartment |
| *apartment rent* [€ / month] | Non-spatial | Continuous | Monthly rent (outcome variable) |

*Observations: 984 apartments*
[a] U*rban district* is modeled as 2D area that encompasses several buildings. Hence, it describes a 2D relation between buildings.

*Method and model selection.* Since we are uncertain about the suitability of prediction methods, we evaluate a broad, heterogeneous candidate set of linear, non-linear, and spatial methods. Conventional prediction methods are represented by an ordinary least squares (OLS) regression, an OLS regression with elastic net regularization, a random forest, and a support vector regression. Among the spatial prediction algorithms, we choose a spatial error model, which models distance-based associations between observations in the error term[3].

---

[2] Descriptive statistics are given in the online appendix.

*Model evaluation.* To evaluate the benefits of modeling in 3D, we compare the full 3D model to more naive models relying on 2D and 1D feature types. The performance measure is the accuracy (measured as root-mean-square error, RMSE) in predicting apartment rents on a hold-out data set. The results in Table 3 demonstrate that the 3D model outperforms its 2D and 1D counterparts regardless of the prediction method. The increase in predictive accuracy of the 3D models in comparison to the 2D models ranges between 2 and 18 percent, depending on the prediction method. Overall, the random forest performs best. In addition, considering potential spatial correlation with the spatial error model does not provide higher accuracy than the non-spatial methods.

**Table 3 Performance of 3D, 2D and 1D models for predicting rent prices**

| Model | number of rooms | apartment size | district | latitude | longitude | elevation | volume | SEM | RF | SVR | OLS | OLSNet |
|---|---|---|---|---|---|---|---|---|---|---|---|---|
| Intercept | | | | | | | | n.a.† | 645.15† | 645.15† | 645.15† | 645.15† |
| 1D | ✓ | | | | | | | n.a. | 441.77 | 442.79 | 428.98† | 428.98† |
| 2D | ✓ | ✓ | ✓ | ✓ | ✓ | | | 361.09 | 259.10† | 291.76 | 287.72 | 289.45 |
| **3D** | ✓ | ✓ | ✓ | ✓ | ✓ | ✓ | ✓ | **305.62** | **254.49†** | **283.50** | **281.02** | **282.00** |

Observations: 984 buildings †best prediction method
Reported performance measure: root mean squared error (RMSE)
Dependent variable: *apartment rent*
SEM = spatial error model
RF = random forest
SVR = support vector regression
OLS = ordinary least squares regression
OLSNet = OLS with elastic net regularization

## 3.3 Showcase: Adoption of Photovoltaic Systems

Photovoltaic (PV) systems are the most common type of solar panel used for renewable electricity generation. Therefore, the adoption of PV systems by consumers is an important issue for policymakers in cities and countries that want to transition to more sustainable energy production and smart grid solutions. While some of the antecedents of technology adoption are well understood on a general level,

---
[3] The mathematical description of the spatial error model is given in the online appendix

for instance perceived monetary costs and benefits (Venkatesh et al. 2012), the translation of these general factors to the specific setting of a particular technology is not necessarily trivial. In the case of photovoltaic (PV) systems, monetary benefits are directly related to the available amount of solar irradiance that the systems can convert into electricity (and that can subsequently be utilized or sold by the owner of the system). Since 3D building geometries allow the incoming solar irradiance to be computed by considering the shading from nearby buildings or hills, 3D analytics should enable a more precise estimation of monetary benefits and the adoption of PV systems. We therefore utilize 3D analytics to design an application that learns PV adoption patterns from several municipal districts and predicts adoption—one in an urban region and one in a rural and hilly area.

*Study design and data collection.* As the incoming solar irradiance at the surface of a PV system depends on the roof characteristics (e.g. its inclination, orientation, and size), we model the building entities in 3D. We also model relations in 3D in order to detect potential shading from neighboring buildings or hills, which can strongly influence the profitability of PV systems.

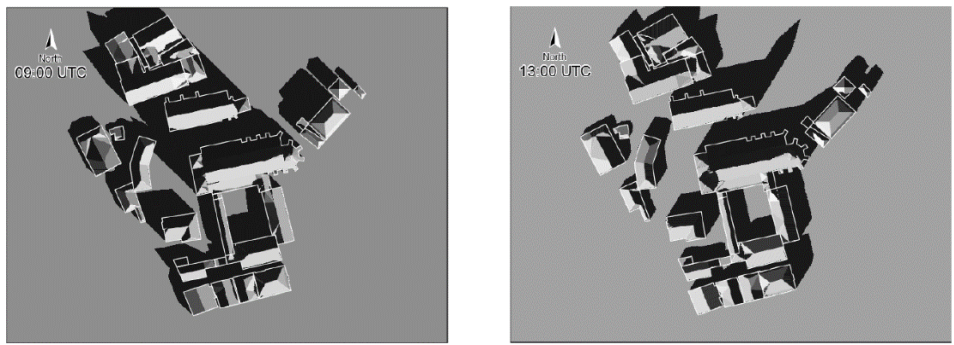

**Figure 2 Shadows on buildings during morning hours (9 am, left panel) and afternoons (1 pm, right panel) on a sample day, as determined by the computational model**

*Feature engineering.* We estimate such shading effects with a model of solar irradiance that computes the sun position at the study areas over one calendar year, taking into account historical weather conditions such as cloud coverage. Figure 2 illustrates this computation for two points in time, one in the morning (left panel) and one in the afternoon (right panel)[4]. The computation results in the feature *PV potential*, which reflects the annual electricity that a PV system could generate on a particular roof surface. Apart from this conceptually complex 3D relation, we describe the urban environment in terms of the 2D relations *municipality, neighborhood*, and *building density*, which have proven to relate to PV adoption, and the 2D geocoordinates of the buildings, *latitude* and *longitude*. To determine features for the entities, we follow the requirements of the application and focus on capturing the roof characteristics of the buildings. In detail, we derive the features *roof orientation, roof inclination, roof surface,* and *roof type*. Overall, feature engineering results in the variables given in Table 4.

---

[4]The computational procedure is described in the online appendix, including a video that illustrates the shading over the course of the day.

**Table 4 Features and types in PV showcase[5]**

| Feature | Type | Value range | Explanation |
|---|---|---|---|
| PV potential [h/a] | 3D relation | Continuous | Potential yearly electricity generation on roof surface |
| municipality | 2D relation[a] | {A, B} | Municipality in which the observation is situated. |
| neighborhood | 2D relation[a] | {historic, residential, commercial, industrial, mixed} | Primary settlement function of the neighborhood. Settlement function affects PV adoption (Dewald and Truffer 2012); for instance, regulations can prohibit PV systems in historic neighborhoods. |
| building density | 2D relation | {low, medium, high} | Building density (per square kilometer), classified as either low, medium or high. An increase in building density leads to a decrease in PV adoption (Graziano and Gillingham 2014). |
| latitude | 2D relation | Continuous | Geocoordinates of the roof surface |
| longitude | | Continuous | |
| roof orientation [°] | 3D entity | Continuous | Cardinal direction of the roof surface; the optimal orientation is south (180°); suitability for PV systems declines gradually for western and eastern orientations. |
| roof inclination [°] | 3D entity | Continuous | Inclination of the roof surface; ranges between 0° for flat roofs and up to 90° for steep surfaces. Optimal angle for PV systems is 30°. |
| roof surface [$m^2$] | 3D entity[b] | Continuous | The surface size of the roof |
| roof type | 3D entity | {flat, A frame, other} | Roof types affect the construction effort for racks that hold PV systems. |
| building function | Non-spatial entity | {main building, outbuilding, housing, business, public, other} | Primary function of the building (outbuildings comprise barns, sheds, and garages) |
| PV system | 1D entity | {present, absent} | Outcome variable; indicates whether a PV system is installed on the roof surface. |

[a] *municipality* and *neighborhood* are modeled as 2D areas that encompass several buildings. Hence, the two features describe a 2D relation between buildings.
[b] Since roof surfaces may be tilted, determining their surface area requires a 3D entity.

***Method and model selection.*** We predict the presence (or absence) of PV systems on roof surfaces. Since this is a classification task, we evaluate a random forest, a support vector machine, a logistic regression, and a logistic regression with elastic net regularization (among non-spatial methods), as well as a logistic spatial error model. To account for the heavy imbalance of the data (PV systems are only observed on two percent of the roof surfaces), we additionally consider a random forest with random oversampling examples of the minority class (Menardi and Torelli 2014) during model training, a prediction method for imbalanced datasets.

***Model evaluation.*** We again compare the predictive performance (measured as the area under the curve (AUC) of the receiver operating characteristic) of the 3D model to simpler models relying on features generated from 2D and 1D entities and relations. The prediction results in Table 5 show that the 3D model considerably outperforms the naiver benchmark models, regardless of the prediction algorithm. The increase in predictive performance of the 3D model ranges between 23 and 73 percent in comparison to the second-best 2D or 1D model. In addition, the spatial error model (SEM) outperforms all conventional prediction algorithms.

---

[5] Descriptive statistics are given in the online appendix.

**Table 5 Performance of 3D, 2D, and 1D models for predicting the presence of PV systems**

| Model | building function | municipality | neighborhood | building density | latitude | longitude | roof type | roof orientation | roof inclination | roof surface | PV potential | SEM | RF | RF ROSE | SVM | Logit | LgNet |
|---|---|---|---|---|---|---|---|---|---|---|---|---|---|---|---|---|---|
| Intercept | | | | | | | | | | | | n.a. | 0.5000 | 0.5000 | 0.5000 | 0.5000 | 0.5000 |
| 1D | ✓ | | | | | | | | | | | n.a. | 0.5000 | 0.5991† | 0.5000 | 0.5991† | 0.5991† |
| 2D | ✓ | ✓ | ✓ | ✓ | ✓ | ✓ | | | | | | 0.7091† | 0.6015 | 0.6904 | 0.5901 | 0.6896 | 0.7029 |
| **3D** | ✓ | ✓ | ✓ | ✓ | ✓ | ✓ | ✓ | ✓ | ✓ | ✓ | ✓ | **0.8902†** | **0.8596** | **0.7850** | **0.7462** | **0.7879** | **0.8791** |

Observations: 16,398 roof surfaces †best prediction method
Dependent variable: *PV system*
Reported performance measure: area under the receiver operating characteristic curve (AUC)
SEM = spatial error model (logistic)
RF = random forest
RF ROSE = random forest with random oversampling examples
SVM = support vector machine
logit = logistic regression
LgNet = logistic regression with elastic net regularization
For the linear methods (SEM, Logit, LgNet), non-monotonous features (roof orientation and roof inclination) were transformed monotonous relations.

While the previous two showcases demonstrated the benefits of 3D analytics in two applications in the urban environment, we next present modeling guidelines for 3D analytics problems in more general applications.

**3.4 Modeling Guidelines for 3D Analytics**

3D analytics projects add specific requirements in terms of resources and knowledge in comparison to conventional analytics and/or 2D spatial analytics projects. In order to arrive at modeling guidelines for 3D analytics, we reviewed general analytics guidelines (Shearer 2000, Fayyad et al. 1996, Shmueli and Koppius 2011) and identified challenges when instantiating them for cases of 3D data. We then derived the modeling guidelines by answering the identified challenges using our insights from the showcases. Challenges and corresponding guidelines are summarized in Figure 3 and explained in the following.

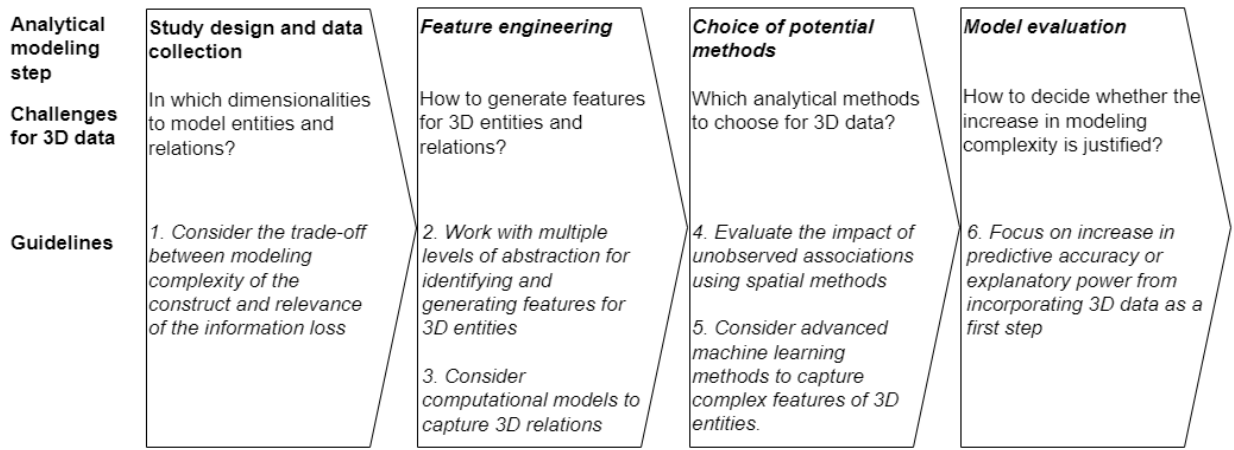

Figure 3 Modeling Challenges and Guidelines for 3D Analytics

### 3.4.1 Study design and data collection: In which dimensionalities should entities and relations be modeled?

During study design and data collection, researchers should select interesting data subsets and variables relying on theory, domain knowledge, and exploratory analysis (Shmueli, 2010; Shearer, 2000; Fayyad et al., 1996). In the 3D setting, the selection of data and variables requires researchers to decide which modeling dimensionalities for entities and relations are required. As previously illustrated, applications do not model all entities and relations of the physical world in three dimensions. Even for the same entity or relation, adequate modeling dimensionalities may vary depending on the application. The challenge is then, *how to decide on the modeling dimensionality of an entity or a relation?* Based on the insights from our showcases, we recommend the following two-step process to making an informed decision.

Entities and relations serve to operationalize different constructs. Therefore, in the first step, the relevance of the construct for the application as well as the relevance of the information loss when modeling the construct in lower dimensionalities should be evaluated. This evaluation has to be based on prior theoretical knowledge in the domain. In the PV showcase, for example, we used 3D building entities to model the construct PV potential. The PV potential is one of the most important determinants of PV adoption since it directly determines a PV system's electricity generation and thus the return-on-investment. Using 2D building entities, i.e. just building footprints, does not allow for a precise measurement of the PV potential because roof inclination and orientation as well as shading from neighboring buildings cannot be captured. The information loss from using 2D footprints instead of 3D building models would thus be so large that the construct 'solar irradiance' would no longer be measurable in a defensible manner. In the real estate case, we made a converse decision concerning the modeling of neighborhood characteristics. As stated by hedonic pricing theory, the distance to amenities considerably influences real estate prices, in addition to the characteristics of the object itself. In our showcase, we approximated the environment using 2D relations (such as the neighborhood, as well as the latitudinal and longitudinal location). Adding the third dimension, i.e. the elevation (above sea level) of the building, would not have considerably altered the overall distance (except in extreme cases where municipalities are situated in a hilly area with very strong gradients).

In the second step, the information loss has to be traded-off against modeling complexity. When modeling in 3D, major sources of complexity are computation, methodological challenges, and data collection. Computational resources for 3D analytics are demanding in several aspects. First, in terms of data storage, 3D data sets are much larger than their 2D counterparts. For example, the data size of 3D city models such as the ones used in the PV and real estate showcases lies in the range of hundreds of gigabytes for large cities and regions, whereas the size of 2D models remains in the range of megabytes. Second, computational effort for data processing is high. Algorithmic models for feature engineering, such as the irradiance computation in the PV showcase and other algorithms (as summarized later in Table 7), are computationally expensive and therefore often require data splitting and sequential processing. Third, data visualization is also computationally expensive and can frequently be carried out only for a limited amount of observations. Beyond computation, working with 3D data also requires interdisciplinary methodological and technical competencies. Many applications rely on advanced geometric and physical models so that advanced knowledge of physics and engineering is required, e.g. to make sense of inertial measurements from wearable sensors or to calibrate a PV potential analysis. Additionally, not all 2D GIS systems or software packages can be used with 3D data. Many of their 3D counterparts are still in developmental stages, therefore advanced technical knowledge is likely needed. As a final decision criterion, data collection effort and cost must be considered. These depend on the data collection technology used: Whereas 3D cameras and wearables are becoming commodities, collecting data using planes & LiDAR is much more expensive[6]. Moreover, there are often also large secondary data sets publicly available, particularly of 3D urban environments, that offer an alternative.

Based on the above criteria, the decision about adequate modeling dimensionalities can be made. To summarize, we recommend that researchers:

<u>*Guideline 1:*</u> *Consider the trade-off between the modeling complexity of the construct and the relevance of the information loss for the application.*

### 3.4.2 Feature Engineering: How should features for 3D entities be generated?

In general, features should be identified and engineered based on theory or empirical association with the dependent variable (Shmueli and Koppius 2011). With 3D data, geometries of 3D entities have many more characteristics than 2D or 1D entities. For example, there are many more options for characterizing a 3D building entity (e.g. height, number of floors, and number and position of windows) than a 2D building footprint, where the options are limited to the size and the shape (e.g. breadth and width) of the footprint. For 3D analytics, the challenge is to decide *how features can be generated for 3D entities and 3D relations.* First, the large number of alternatives makes it more difficult to *identify* informative features. Second, the more complex geometries also make it more difficult to *generate* the features.

We can derive the following insights regarding this challenge from our showcases. In the PV case, on the one hand, for certain constructs we chose features with a low level of aggregation, such as *roof orientation* and *roof inclination*. Since these low-level features directly relate to the physical laws that determine the incoming solar irradiance, they are useful. On the other hand, for other constructs, we chose features with higher levels of aggregation, i.e. features that consider geometries of several parts of the building. An example is the feature *roof type*, which summarizes how several individual roof surfaces are connected to form, for example, an A-shaped roof. The high-level feature *roof type* allowed us to include information about economic aspects of PV adoption into our model, since the cost and the usable area for mounting a PV system vary between roof types. In the real estate case, we also identified features with both low and high levels of aggregation as containing relevant information about the value of an apartment. For instance, the low-level feature *apartment size* (a 2D entity) contains the size of the apartment in square meters and can be derived from apartment footprints. In addition, the high-level feature *building volume* contains information about the overall number of apartment units in the building. An apartment in a building with fewer units generally has a higher value than a comparable one in a building with several units.

---

[6] For an overview of 3D data sources and data capturing technologies, including advantages and disadvantages as well as common modeling libraries, we refer to the online appendix.

While a combination of features with both low and high levels of aggregation proved valuable for our showcases, such an approach is also helpful in other applications with other entities, such as sports analytics. In basketball, for example, the player's body posture is a key determinant of shooting performance. The body posture can be described with a large number of features, such as the geometries of individual bones or the angles of joints. Among all potential geometries, it is reasonable to model the arm of the shooter with a low-level of aggregation, such as the angles of the elbow and the wrist, and to include higher-level features to describe the overall balance of the player during shot release (cf. Felsen and Lucey 2017). Table 6 illustrates levels of aggregation for features of *humans* and *buildings*—entities that are part of a large number of applications in IS research. In summary, we suggest that researchers:

<u>*Guideline 2:*</u> *Work with multiple levels of aggregation for identifying and generating features for 3D entities.*

**Table 6 Features for human and building entities ordered by aggregation levels**

| Level of aggregation | Human | Building |
|---|---|---|
| **Low** | Bones | Walls, roof surface |
| | Angles of joints | Room |
| | | Apartment |
| | Limbs (arms, legs, etc.) | Building |
| | | Street |
| **High** | Overall body posture | City |

### 3.4.3 Feature Engineering: How should features for 3D relations be generated?

In comparison to 3D entities, the complexity in feature generation is even more pronounced for 3D relations. For distance-based relations, distance needs to be computed considering all three dimensions. Moreover, a large number of relations are not purely distance-related, such as line-of-sight and noise-,

heat-, light-, and WiFi-propagation. These relations cannot be approximated satisfactorily with conventional methods, such as distance matrices.

The PV showcase illustrates that the effect of non-distance-based 3D relations, such as solar irradiance and shading, can vary strongly even for locations that are very close. For example, rooftops of the same building but with opposite orientations have a vastly different PV potential. To capture this complex 3D relation at a very granular level, we needed to use a computational model.

Table 7 Types of 3D relations and quantification approaches

| Type of 3D relation | Quantification approach | Algorithmic Implementations | Application examples |
|---|---|---|---|
| **Distance-based** | | | |
| Manhattan distance | Formula(x,y,z) | E.g. spatial regression methods | Applied in numerous domains |
| Euclidean distance | Formula(x,y,z) | E.g. spatial regression methods | Applied in numerous domains |
| **Not primarily distance-based** | | | |
| Walking distance | Computational model | Shortest path algorithms | Walking distance to determine trajectories in a mall (Ghose et al. 2019) |
| Solar irradiance | Computational model | Corripio (2021), Yang et al. (2018) | Solar irradiance to analyze PV potential (this paper) |
| Line-of-sight | Computational model | Domain-dependent, see Biljecki et al. (2015) for an overview | Line-of-sight to detect TV viewer's attention (Leroy et al. 2013) |
| Sound propagation | Computational model | Stoter et al. (2008) | Sound propagation to analyze traffic noise in cities (Stoter et al. 2008) |
| Radio-wave propagation | Computational model | He et al. (2019) | Indoor positioning methods, design of urban communication infrastructures (WiFi, cellular) |
| Visibility | Computational model | Bartie et al. (2010) | Visibility analysis to determine sea view from apartment (Dong et al. 2015, Yu et al. 2007) |

Such complex 3D relations are not limited to our showcase on PV adoption, but are a common phenomenon when working with 3D data. Table 7 illustrates types of 3D relations, quantification approaches, algorithmic implementations, and associated applications. For the group of relations that are not primarily distance-based, such as solar irradiance, line-of-sight, sound propagation and visibility, researchers should resort to computational models. Even for some distance-based relations, such as walking trajectories, computational models are required, as walking trajectories are not as easily calculated using a standard formula like Euclidian distance. In sum, the examples highlight that computational models are an indispensable tool for feature generation for 3D analytics. We therefore suggest that researchers:

*Guideline 3:* *Consider computational models to capture 3D relations.*

**3.4.4    Method and Model selection: Which analytical methods should be chosen for 3D data?**

During the method and model selection phase, the researcher needs to fit a set of analytical methods and models to the data and evaluate their suitability for the task at hand. The standard approach is to evaluate a set of analytical methods and eventually choose the model with the highest explanatory or predictive power (Shmueli and Koppius 2011, Shearer 2000). Some analytical methods have rather general applications, such as random forests or support vector machines, while others perform particularly well with specific data sources, such as neural networks with image data and time-series methods for time-series data. For 3D analytics, the challenge is to decide *which methods are suitable for analyzing 3D data.* While spatial regression methods are tailored towards capturing spatial associations as a function of distance between observations, it is unclear whether these methods also perform well with 3D data. 3D relations, such as line-of-sight, can cause associations that are not determined by distance alone and whether distance-based methods can sufficiently capture such relations likely depends on the specific application.

As a result of this uncertainty, in the showcases, we evaluated a broad set of linear and non-linear methods. However, we expected that there might be additional effects from unobserved spatial associations between the observations in our datasets, resulting from relations for which we did not generate features. In the real estate case, hedonic pricing theory led us to believe that characteristics of the urban environment, such as the proximity to amenities, may cause rent prices to vary locally. Similarly, in the PV case, we expected that peer effects (Bollinger and Gillingham 2012) could result in PV systems being geographically clustered in our study area. Because of such unobserved effects, we decided to also fit a spatial error model to the data in both showcases. While the spatial error model did not provide superior performance in the real estate case, accounting for spatial associations in the solar case did indeed lead to an increase in predictive performance. Since such spatial dependencies from unobserved relations are common in 3D settings, we suggest that researchers:

<u>Guideline 4:</u> *Evaluate the impact of unobserved associations using spatial methods.*

Spatial statistics has provided a large number of methods for modeling spatial associations. These methods can also be applied with 3D data; however, the 3D setting allows for a greater freedom in modeling these associations. First, while it is common for spatial analytics applications to model associations in the horizontal plane—i.e. as a distance-dependent *bivariate* function $d(x, y)$, 3D data allows one to take into account associations in three dimensions $d(x, y, z)$. Second, the nature of associations may also be anisotropic, i.e. they vary in different directions. For example, in a study on organizational knowledge sharing, Reagans (2011) discovered that vertical separation of workplaces (i.e. workplaces located on different floors) reduces communication frequency among employees much more severely than horizontal separation. If such effects are expected, researchers should choose spatial methods that can take anisotropy into account.

Apart from hidden relations between entities, the entities themselves may motivate the use of advanced machine learning methods, such as geometric deep learning (Bronstein et al. 2017, Griffiths and Boehm 2019). These methods do not require explicit feature engineering, instead they directly take 3D entities as input in the format of 3D point clouds or 3D polygon meshes (as presented earlier in Table 1) and generate feature representations internally. The use of such methods is advantageous if feature engineering from 3D geometries cannot be standardized, for example if heterogeneous geometries represent the same feature value. This is the case in human-computer interaction when an application requires the identification of body parts (e.g. hands) from different persons that may additionally be in captured in different body postures. Another example is the derivation of abstract features from geometries, such as architectural styles from building facades. In these and related tasks, explicit feature engineering cannot be standardized, since the connection between geometries and the features is complex. In such cases, we suggest that researchers:

*Guideline 5:* *Consider advanced machine learning methods to capture complex features of 3D entities.*

### 3.4.5 Model Evaluation: How to decide whether the increase in modeling complexity is justified?

The final stage involves reporting performance measures and deploying the model (Fayyad et al. 1996, Shearer 2000). Since 3D data makes analytical models conceptually and computationally more complex, the key challenge at this stage is *how to decide whether this added complexity is justified*.

The PV and real estate examples showed that comparing the 3D models with models relying on lower dimensionalities was a suitable benchmark for isolating the benefits of 3D data. Whereas we achieved a large increase in predictive accuracy (23–73 percent) for the 3D model in the PV showcase, the increase in predictive accuracy in the real estate showcase was comparatively small (2–18 percent). Ultimately, deciding whether these increases in model performance justify the additional modeling effort depends on the increase in theoretical understanding or the practical value. The practical benefits of a PV model, for instance, could be evaluated in an application for targeted marketing that assists PV system vendors in identifying customers likely to be interested in such systems.

Using lower-dimensional models to isolate the benefits of 3D data is also a common approach beyond our showcases. For instance, Ghose et al. (2019) outline the advantages of their model for location-based advertisement using 3D trajectories in comparison to simpler models that do not use this 3D relation. While the authors quantify the benefits in terms of increased advertising effectiveness, this may not be possible in all settings. In summary, we suggest that researchers:

*Guideline 6:* *Focus on the increase in predictive performance or explanatory power from incorporating 3D data as a first step.*

In summary, the six guidelines point towards crucial aspects and trade-offs for 3D analytics implementations. It remains to note that the guidelines should not be interpreted as strict decision rules, researchers are rather required to add their own domain knowledge and judgement when operationalizing them. Nevertheless, the guidelines provide researchers with a structured approach—from study design and data collection, to feature engineering, methods and model selection, and model evaluation—to design their 3D analytics applications.

## 4    Benefits of 3D Analytics for IS Research

After focusing on applying 3D analytics, including showcases and modelling guidelines, we now look at the benefits for IS research offered by 3D analytics. On the one hand, 3D analytics helps to address several research opportunities in different contextual areas of IS research (described in Section 4.1); on the other hand, 3D analytics can support researchers in conducting several tasks common to IS research projects (focus of Section 4.2).

## 4.1 Research Opportunities

In order to identify research opportunities to be addressed with 3D analytics, we reviewed open research questions described in seminal papers and review articles in key contextual areas of IS research, such as healthcare IS, mobile commerce, human-computer interaction and others.[7] Thereafter, we analyzed whether and how 3D analytics can help addressing them.

We find that 3D analytics helps to advance IS research particularly in three major research traditions: (1) 3D analytics can support behavioral IS research by increasing our understanding of the behavior of humans when interacting with IT; (2) 3D analytics enables improved decision support in several areas, and (3) 3D analytics provides opportunities for designing novel and/or enhanced information systems. An overview of the identified research opportunities for 3D analytics is given in Table 8. In the following, we explain the opportunities individually for each contextual research area.

**Table 8 Research opportunities to be addressed with 3D analytics**

| | Behavioral IS research | Decision support | IS design |
|---|---|---|---|
| **Healthcare IS** | Understand the effect of personalization of health IS enabled by wearables on adoption and use, on health outcomes, as well as the doctor-patient relationship | Improve healthcare predictive modeling using on 3D representations of the human body<br><br>Use 3D analytics to better understand causes of injuries | Design real-time health monitoring services |
| **Human-computer interaction** | Understand the relevance of the spatial configuration in spatially distributed IS, such as cyberphysical systems, human-robot interaction, AR & VR applications | | Design interactive TV applications<br><br>Improve services provided in spatial IS, such as smart homes |
| **Mobile commerce** | Understand the relevance of the hypercontext in mobile, location-based marketing | Decision support for store layout planning | Improve location-based advertisement applications using hyperlocal targeting |
| **Green IS & energy informatics** | Understand antecedents of the adoption of sustainable technologies (e.g. smart homes, photovoltaic systems) | Enhance decision support systems for the planning of energy infrastructure<br><br>Build government support systems for evidence-based energy policy design | Develop methods to generalize scarce sensor data<br><br>Detect target customers for energy-related products based on 3D building characteristics.<br><br>Provide benchmarks for resource consumption based on 3D building characteristics. |
| **Smart cities & public sector digitalization** | | Use 3D digital twins of cities for planning neighborhoods for a higher quality of life<br><br>Improve demand estimations for location-based services using neighborhood characteristics (e.g. building volumes) | Increase the speed and accuracy of administrative processes (e.g. tax collection, PV system registration)<br><br>Use 3D building geometries to provide digital real-estate valuation services |
| **Sports analytics** | Understand coordination, risk-taking, and errors in teams | Create DSS for coaching in real-time | Applications for performance monitoring and training |
| **Summary** | **3D analytics helps to understand the behavior of humans when interacting with IT** | **3D analytics helps to derive insights for improved decision making** | **3D analytics enables the design of novel or enhanced information systems** |

---

[7] References to the most relevant literature sources are provided in Appendix F.

### 4.1.1 3D Analytics for Healthcare IS

In recent years, healthcare IS have had considerable success in providing better, more cost-efficient and more patient-centric healthcare (Bardhan et al. 2020). 3D data of the human body collected by wearable sensors and/or 3D cameras can further this success in two ways. First, it can enable a more refined theoretical understanding of how injuries arise and what can be done to prevent them. In the past, researchers often had to reduce the human body to easily measurable properties such as height and weight or the measures of various limbs, to use as control variables in predicting injury risk (e.g. differences in leg length can affect the risk of injuries among runners (Bennell et al. 1996)). At the same time, these variables are proxies for underlying causes of injuries such as left-right balance or where the body's center of gravity is in relation to its base of support. Using 3D analytics allows one to measure these underlying causes with much more precision, thus leading to a more refined understanding of injuries.

Second, the information provided by wearable 3D sensors can also provide the basis for the design of advanced health DSS that provide feedback in real-time. An example is the prediction and prevention of falls among the elderly. The majority of such fall prevention systems are currently still relying on lab experiments and ex-post analysis (Howcroft et al. 2013). However, when used in real-time with mobile applications or environmental feedback, they can prevent falling as well as subsequent injuries and can thus have significant impact on health and healthcare cost. A similar example is the use of inertial sensors to determine 3D aspects of human movement. In a recent lab experiment, Bramah et al. (2018) examine the differences in 3D kinematics between healthy and injured runners and find that injured runners have a higher forward lean of the trunk and exhibit a larger vertical movement of the hips. While this application is currently still at the stage of controlled lab experiments, the findings can eventually be used to design information systems in the field by embedding sensors into clothing to measure these 3D aspects in real time and give feedback to runners to develop a healthy running style.

Moreover, healthcare predictive modeling has brought significant progress for evidence-based treatments, such as the prediction of adverse events and risk analytics related to diseases (Bardhan et al. 2020, Fichman et al. 2011). For example, deep learning algorithms applied to X-ray images have proven successful in supporting disease detection. In addition to such image data, 3D representations of organs and limbs from computer tomographs combined 3D analytics can further increase the accuracy of such predictive models for disease detection. First successes have already been achieved with 3D deep learning methods in the area of lung cancer detection (cf. Perez and Arbelaez 2020), but there are many more applications to be explored.

Eventually, the mentioned applications will lead to more personalized and evidence-based health services. This provides important follow-up research questions, such as whether these advantages will lead to an increase in the adoption of health IS and how they alter the doctor-patient relationship (Agarwal et al. 2010).

### 4.1.2 3D Analytics for Human-Computer Interaction

In human-computer interaction (HCI) research, IS scholars are concerned with ways humans interact with information, technologies and tasks (Zhang et al. 2009).

Using 3D data captured by depth cameras and wearables, combined with 3D analytics, provides new opportunities to HCI research, particularly to better understand the relevance of the spatial configuration between user and IT artefact. For instance, in 'second screen' (or multi-screen) applications (cf. Hinz et al. 2016), 3D analytics can help to observe the head pose of the user and thereby understand how the user divides her attention, for instance, between a TV screen and a tablet. Similarly, in human-robot interaction, 3D analytics allows robots not only to notice when we are paying attention (van der Pol et al. 2011), but also to observe our gestures and to draw associated conclusions on the effectiveness of the interaction.

Overall, 3D analytics can thereby improve digital services, such as interactive TV applications or the generation of user profiles based on our attention preferences. Moreover, it can inform the design of spatially distributed cyberphysical systems, such as smart homes. Beyond improving applications, a better understanding of such spatial interactions can also help to refine related theories, e.g. on distributed cognition and media richness.

### 4.1.3 3D Analytics for Mobile Commerce

In mobile location-based commerce, smartphone ads and push messages are tailored to the current location and surrounding environment of a specific user. Recently, research has been focusing on the impact of the hypercontext, i.e. the immediate environment of the user, on the effectiveness of marketing measures. Fang et al. (2015) demonstrate that mobile ads for movies are more effective when sent to consumers in close proximity (500 meters) to the theater. Ghose et al. (2019) demonstrate that targeting ads according to users' trajectory patterns through a large shopping mall has a positive impact on mobile marketing effectiveness.

3D analytics provides opportunities for increasing our understanding of the relevance of the hypercontext for mobile location-based marketing. For example, Rook (1987) found in a qualitative survey that an important determinant of buying behavior is line-of-sight. However, the question of how line-of-sight impacts the effectiveness of location-based advertising has not yet been answered. By modeling line-of-sight in shopping malls and in the interior of stores—including viewsheds from shelves and other impediments, we can test and refine Rook's theory on the impact of line-of-sight on buying behavior quantitatively. Moreover, Fang et al. (2015) demonstrate that location-based advertisements have positive immediate and delayed effects. 3D analytics provides opportunities to test whether the delayed effects are stronger for people having line-of-sight when targeted, as perceiving the targeted store visually may increase the memory effect and therefore purchasing plans. 3D cameras, 3D indoor models, and analytics allow for the automation of data collection and the analysis of these and similar theoretical questions using large data samples.

A related research opportunity focuses on evaluating the effect of crowding and mobile immersion on both spending and advertisement effectiveness. It is known that crowding in stores attracts shoppers to the crowded areas but at the same time reduces spending in these areas (Hui et al. 2009). Similarly, consumers respond more often to mobile ads when in a crowded vs. non-crowded environment, which may be due to mobile immersion (Andrews et al. 2016). 3D analytics allows for a test of whether the 3D configuration, such as the height of floors and shelves, affects perceived crowdedness and marketing effectiveness.

The findings regarding the previous questions may not only improve the theoretical precision of location-based advertising research but also improve location-based marketing applications. On the one hand, systems to target ads according to shopper's locations (cf. Ghose et al. 2019) can be refined based on whether they have line-of-sight to stores or even products within stores. On the other hand, the same mechanism may also be used by marketers to take advantage of unplanned spending (cf. Hui et al. 2013). By sending mobile ads to shoppers, their trajectories may be altered so that they are visually attracted to more stores or products. Finally, the aforementioned theoretical insights can also be incorporated into decision support systems for the design of the physical layout of stores and shops (cf. Vrechopoulos et al. 2004, Sevilla and Townsend 2016).

### 4.1.4 3D Analytics for Green IS and Energy Informatics

Researchers in Energy Informatics and Green IS aim to reduce the consumption of scarce environmental resources and associated emissions (Watson et al. 2010, Goebel et al. 2014). They develop IS for increasing energy efficiency, reducing electricity consumption for heating/cooling, increasing the share of renewable energy production, reducing noise and pollution from traffic, and others.

The design of new systems and services in this area is, however, inhibited by the lack of sensor data (Watson et al. 2010). For example, smart meters are only installed at a minority of buildings and underly privacy restrictions, which is a challenge for the planning of energy infrastructures and other applications. Here, 3D analytics offers large potential to reduce the required amount of sensor data. Thanks to the fact that many processes, such as energy use, noise and heat diffusion, relate to building geometries, 3D city models can be used to generalize from the few sensor measurements available.

3D city models combined with analytics can also improve a variety of energy-related decision support systems and digital services. For example, more precise estimates of resource consumption and pollution based on 3D data can advance government support systems to derive effective energy policies. Moreover, 3D analytics enables detecting target customers for energy-related products (e.g. photovoltaic (PV) systems and building retrofits), and improving benchmarks for building resource consumption (cf. Loock et al. 2013), including water, heat, and electricity.

### 4.1.5 3D Analytics for Smart Cities and Public Sector Digitalization

In recent years, cities and metropolitan areas have been undergoing a fundamental transformation as a result of digitalization (Chourabi et al. 2012). 3D building and cities models are a type of data that is increasingly becoming available as part of this process. The usage of such 3D digital twins of buildings is currently receiving attention in the construction sector and in engineering-related applications

(Biljecki et al. 2015). The association of this type of data with socio-economic processes, however, offers large potentials for research on smart cities and the digitalization of the public sector.

First, 3D city models can be used to provide decision support for planning and operating cities for an increased quality of life. For example, to foster communication and social cohesion, 3D models can be used to plan neighborhoods to increase visual and personal contact between residents. Similarly, 3D models can help to design neighborhoods for increasing public security by avoiding the construction of secluded and insecure areas. Another set of potential applications lies in the facilitation of administrative processes. Here, first research has already proposed automating the collection of taxes or the registration of photovoltaic systems based on 3D building data (Ali et al. 2018, Rausch et al. 2020).

Second, the association between building geometries and socio-economic activity can also be leveraged to transform existing digital services and to create new ones. A crucial challenge for many digitally-enabled services in the urban environment, such as mobility services, is the estimation of the spatial distribution of consumer demand, which is often based on points-of-interest and related data sources (Brandt et al. 2021, Wang et al. 2020). Here, 3D volumes of buildings may be a better proxy as they closely relate to the number of persons working or living in them. Another application, for which 3D building models can be particularly valuable, is real estate (as we demonstrate in our showcase). The real estate industry is particularly disrupted by the digitalization with real estate agents being increasingly replaced by property technology (proptech) companies that provide real estate valuations based on spatial big data (Braesemann and Baum 2020). Overall, these example are only indicative of how 3D analytics can advance decision support systems and IS design in the urban environment.

### 4.1.6 3D Sports Analytics

In sports, the use of player tracking systems for athletes has a long history (Lucey et al. 2013, Gudmundsson and Horton 2017). Wearables and 3D cameras can not only enhance such systems but also enable several new applications. When considering team sports such as basketball, one application of 3D analytics is to look at the movement and performance of individual players in the team. For instance, the 3D movements of the arm and the head have been shown to affect basketball shot performance (Felsen and Lucey 2017). While still in the stage of prototypes, the findings of such studies may eventually inform the design of IS for coaching by embedding sensors into clothing to measure these 3D aspects in real time and give immediate feedback. Sports teams also have a broader relevance, as they have been used for studying coordination in teams for many years. Examples include studies looking at the role of tacit knowledge (Berman et al. 2002) and social capital (Fonti and Maoret 2016), using a structural perspective on relations between teammates. 3D analytics enables to incorporate more sophisticated aspects, such as shared awareness between teammates, not just by analyzing players' location on the court, but also teammates their field-of-vision (cf. Bourbousson et al. 2015) to identify implicit coordination processes. Such aspects related to situational awareness can be useful to better understand issues such as coordination, risk-taking, and errors in teams in other situations, for instance coordination patterns in emergency response teams (Faraj and Xiao 2006). This in turn could help to design information systems to support coordination in high-pressure situations (Kane and Labianca 2011).

### 4.2 Benefits of 3D Analytics for Research Tasks

Apart from advancing research in the aforementioned contextual areas, 3D analytics can assist in completing tasks common to IS research projects independent of specific application areas. Examples for such generic tasks are the identification of data sources, data collection, modeling as well as experiment design.

To identify these tasks, we reviewed our showcase studies and the research opportunities listed in the previous section and analyzed *how* the benefits from applying 3D analytics materialize. In addition, we compared 3D analytics with other formerly novel types of data and tools, such as NeuroIS (Dimoka et al. 2011), and evaluated whether the benefits associated with these data types apply to 3D analytics, as well. As a result, we identified the following IS research tasks that can be supported using 3D analytics.

#### 4.2.1 3D Data as an Additional, More Direct Data Source

In IS research, data is frequently collected via self-reports or via observational methods (such as case studies or ethnographies). Depending on the application, such data collection may be difficult and, additionally, may result in biases, such as subjectivity bias, social desirability bias and common method bias. For example, research on emotions has suffered due to the subjectivity involved in reliably measuring emotions with self-reports (Dimoka et al. 2011). Another example is location-based marketing, where subjects may not remember whether an advertisement was visible to them or not when surveyed after a field experiment.

For many applications, 3D data can be a more direct data source than self-reports or surveys, reducing the impact of bias. For example, with 3D analytics we can measure the field-of-vision of people exposed to advertisements. We can also record gestures of the human body and relate them to feelings and attitudes. Also, when studying physical objects, e.g. buildings for real estate valuation, it may be more accurate to rely on exact geometric data, since self-reports, such as "spacious terrace" and "nice view", may be biased. For these and related applications, 3D data can be more direct than self-reports and may thus be a more suitable data source.

### 4.2.2 Non-obtrusive Data Collection with 3D Analytics

In field experiments, it is important to collect data in a way that obtrudes the natural behavior of the subject as little as possible or even not at all. Observational research methods, where researchers are present in the field, alter the natural experience and can therefore be problematic.

Using 3D sensors such as wearables or 3D cameras in field experiments permits the collection of data on human behavior in a less obtrusive manner. For example, in a study on impulse buying behavior (which is known to be impacted by line-of-sight (Rook 1987)), data collection via 3D cameras may be more suitable than self-reports or field observations. Similarly, in a study on human-computer interaction with smart home technologies, it may be inadequate to intrude in the private home environment; instead, 3D data collection technologies are a fewer interfering means of data collection.

### 4.2.3 Automation of Data Collection

Traditionally, many phenomena that involve three-dimensional geometric aspects could only be measured by making manual observations in the field. For instance, in a study on real estate value, Helbich et al. (2013) analyzed the effect of panoramic view from apartments on their respective sales prices. For data collection, the authors visited each of the apartments in their data sample to personally assess the quality of the view from the apartment windows. Another application area where data is typically collected manually with observational methods is the design of distributed cyberphysical systems (e.g., smart homes), where researchers often visit the field in order to gather data on the interaction between subjects and devices. In general, due to the large effort, manual data collection limits the sample sizes of such studies.

3D analytics helps to automate data collection and thereby enables studies at a much larger scale. Thanks to airborne LiDAR or stereo photogrammetry, three-dimensional digital twins of entire cities can be generated. With the help of these 3D models, much of the aforementioned data collection can be conducted using computational tools that can derive complex features, such as panoramic views, from the data. Similarly, by installing cameras that capture 3D geometries in the field and by subsequent analytical processing, data on human-computer-interaction can be recorded in an automated fashion. In sum, these examples illustrate how 3D analytics speeds up the data collection process[8]. Thereby, 3D analytics improves the data foundation of studies and improves the statistical backing of findings.

### 4.2.4 New Measurements for IS Constructs

For several popular IS constructs, such as *intentions to use IT* or the *perceived usefulness* in the area of technology acceptance, there are theories in other fields that can inform IS research. For example, it is known from communication science how gestures (such as defensive arm movements, crossed-legs, stepping back) are related to feelings and attitudes (as illustrated in Saitz and Cervenka 2019). Similarly, in ergonomics, there is theory about how body postures facilitate or hinder the use of digital technology (Albin and McLoone 2014).

With 3D sensors (cameras or wearables) and analytics we can measure body postures, gestures, or facial expressions of users working with IT (Molchanov et al. 2016). Thereby, IS researchers can draw on theory from other fields, such as communication science and ergonomics, and create new measurements for important IS constructs. While previously these were mostly measured using questionnaires, 3D data can complement or replace such data sources and thereby provide more accurate and reliable measurements.

---

[8] We explain prominent data capturing technologies for 3D data, including advantages and disadvantages in terms of cost, scalability and precision, as well as common application areas in the appendix.

**4.2.5  Use 3D Models to Approximate Underlying Mechanisms**

In order to test and validate a theory, researchers need to build accurate models of causal mechanisms. This is often challenging, because many real-world phenomena are the results of an interplay of multiple mechanisms. Building a complex model that integrates these mechanisms is difficult: On the one hand, the complexity of the interplay obscures cause-and-effect relations; on the other hand, data is often not available for all inputs. An example is modeling electricity generation of PV systems: A precise physical model of the potential output would depend on system capacity, solar irradiance, temperature, module technology, and other factors. As we demonstrate in our showcase on PV adoption, instead of collecting data on all these inputs, the electricity output can be approximated based on 3D geometries alone using a computational model of solar irradiance combined with a data-driven model relying on roof characteristics. This benefit is not limited to mechanism from physics, with 3D analytics we can also test how e.g. 3D city models relate to socio-economic phenomena, such as the demand for services, or attractiveness and security of neighborhoods. In summary, instead of combining several mechanisms individually in a complex model for which data may be difficult to obtain, approximating them by relying on 3D geometries can be a reasonable alternative.

**4.2.6  New Opportunities for Experimental IS Research**

Apart from the previous advantages for individual tasks in research projects, 3D analytics may also offer new ways of conducting research, particularly lab experiments. Instead of traditional lab experiments in physical labs, 3D data allows for the creation of augmented or virtualized labs. Similar approaches have used environments based on multiplayer online games (Williams et al. 2011). Using 3D analytics (combined with sensors, such as wearables or 3D cameras) the behavior of subjects or their virtual counterparts can be analyzed whilst in these environments (e.g. in a virtual home). This provides advantages, such as saving cost for the physical lab equipment and allowing for better replicability due to a more controlled environment.

In summary, the benefits of 3D analytics—ranging from a more direct data source, to an advantageous means of data collection, to an improved modeling of constructs and mechanisms—are manifold and support several important phases of IS research projects.

# 5  Concluding Remarks

3D analytics represents an exciting opportunity for IS researchers operating at the intersection of human behavior, technology, and the environment. Traditionally, limitations in data availability have forced researchers to make considerable simplifications for modeling the physical environment and humans using technology in this environment. With the advent of sensors in mobile phones and clothing, depth cameras, drones, and other technologies, these limitations are being gradually overcome. Researchers in mobile commerce can now disentangle the effects of visibility of the marketing target from the effect of mere proximity to the target. Researchers in smart cities can now use elevation, view, and shadowing to develop a better understanding of urban technology adoption or to design more effective information systems for public services. Researchers in human-computer interaction can now study the effects of second screening by detecting where people's attention is focused, thus being able to test the effects of mindlessness or mindfulness in more detail. Beyond these areas, researchers can take advantage of 3D analytics to access new data sources, facilitate their data collection and build new models of constructs and mechanisms. Even though operationalizing 3D data adds a layer of complexity to analytical modeling, the given opportunities and task-related benefits together with our modeling guidelines, should motivate and support IS researchers to start their journey into this largely unexplored third spatial dimension.

**References**

Abbasi A, Sarker S, Chiang R (2016) Big Data Research in Information Systems: Toward an Inclusive Research Agenda. *Journal of the Association for Information Systems* 17(2):i–xxxii.

Agarwal R, Gao G, DesRoches C, Jha AK (2010) Research Commentary —The Digital Transformation of Healthcare: Current Status and the Road Ahead. *Information Systems Research* 21(4):796–809.

Aggarwal JK, Xia L (2014) Human activity recognition from 3D data: A review. *Pattern Recognition Letters* 48:70–80.

Albin TJ, McLoone HE (2014) The Effect of Tablet Tilt Angle on Users' Preferences, Postures, and Performance. *Work* 47(2):207–211.

Ali, DA, Deininger, K, Wild, M (2018) *Using Satellite Imagery to Revolutionize Creation of Tax Maps and Local Revenue Collection* (World Bank, Washington, DC).

Andrews M, Luo X, Fang Z, Ghose A (2016) Mobile Ad Effectiveness: Hyper-Contextual Targeting with Crowdedness. *Marketing Science* 35(2):218–233.

Bajracharya A, Reader K, Erban S (2019) User Experience, IoMT, and Healthcare. *THCI*:264–273.

Bardhan I, Chen H, Karahanna E (2020) Connecting systems, data, and people: A multidisciplinary research roadmap for chronic disease management. *MIS Quarterly* 44(1):185–200.

Bartie P, Reitsma F, Kingham S, Mills S (2010) Advancing visibility modelling algorithms for urban environments. *Computers, Environment and Urban Systems* 34(6):518–531.

Bennell KL, Malcolm SA, Thomas SA, Reid SJ, Brukner PD, Ebeling PR, Wark JD (1996) Risk factors for stress fractures in track and field athletes. A twelve-month prospective study. *The American Journal of Sports Medicine* 24(6):810–818.

Berman SL, Down J, Hill CWL (2002) Tacit Knowledge as a Source of Competitive Advantage in the National Basketball Association. *AMJ* 45(1):13–31.

Biljecki F, Stoter J, Ledoux H, Zlatanova S, Çöltekin A (2015) Applications of 3D City Models: State of the Art Review. *IJGI* 4(4):2842–2889.

Bollinger B, Gillingham K (2012) Peer Effects in the Diffusion of Solar Photovoltaic Panels. *Marketing Science* 31(6):900–912.

Bourbousson J, R'Kiouak M, Eccles DW (2015) The Dynamics of Team Coordination: A Social Network Analysis as a Window to Shared Awareness. *European Journal of Work and Organizational Psychology* 24(5):742–760.

Braesemann F, Baum A (2020) PropTech: Turning Real Estate Into a Data-Driven Market? *SSRN Journal*.

Bramah C, Preece SJ, Gill N, Herrington L (2018) Is There a Pathological Gait Associated With Common Soft Tissue Running Injuries? *The American Journal of Sports Medicine* 46(12):3023–3031.

Brandt T, Dlugosch O (2021) Exploratory data science for discovery and ex-ante assessment of operational policies: Insights from vehicle sharing. *Jrnl of Ops Management* 67(3):307–328.

Brandt T, Ketter W, Kolbe LM, Neumann D, Watson RT (2018) Smart Cities and Digitized Urban Management. *Business & Information Systems Engineering* 60(3):193–195.

Brandt T, Wagner S, Neumann D (2021) Prescriptive analytics in public-sector decision-making: A framework and insights from charging infrastructure planning. *European journal of operational research* 291(1):379–393.

Bronstein MM, Bruna J, LeCun Y, Szlam A, Vandergheynst P (2017) Geometric Deep Learning: Going beyond Euclidean data. *IEEE Signal Process. Mag.* 34(4):18–42.

Chourabi H, Nam T, Walker S, Gil-Garcia JR, Mellouli S, Nahon K, Pardo TA, Scholl HJ (2012) Understanding Smart Cities: An Integrative Framework. Sprague RH, ed. *Proceedings of the 45th Annual Hawaii International Conference on System Sciences*: 4-7 January 2012, Maui, Hawaii (IEEE Computer Society, Los Alamitos, Calif.), 2289–2297.

Çöltekin A, Lokka I, Zahner M (2016) On the usability and usefulness of 3D (geo)visualizations - A focus on virtual reality environments. *Int. Arch. Photogramm. Remote Sens. Spatial Inf. Sci.* XLI-B2:387–392.

Corripio JG (2021) insol: Solar Radiation. *R package*, http://CRAN.R-project.org/package=insol.

Cyganek, B, Siebert, JP (2011) *An introduction to 3D computer vision techniques and algorithms* (John Wiley & Sons).

Dewald U, Truffer B (2012) The Local Sources of Market Formation: Explaining Regional Growth Differentials in German Photovoltaic Markets. *European Planning Studies* 20(3):397–420.

Dimoka A, Pavlou PA, Davis FD (2011) Research Commentary —NeuroIS: The Potential of Cognitive Neuroscience for Information Systems Research. *Information Systems Research* 22(4):687–702.

Dong, P, Chen, Q (2017) *LiDAR Remote Sensing and Applications* (CRC Press, Boca Raton, FL : Taylor & Francis, 2018.).

Dong Y, Tang J, Chawla NV, Lou T, Yang Y, Wang B (2015) Inferring Social Status and Rich Club Effects in Enterprise Communication Networks. *PloS one* 10(3):e0119446.

Eves, H (1963) *A Survey of Geometry: Volume One* (Allyn and Bacon, Boston).

Fan C, Zhang C, Yahja A, Mostafavi A (2021) Disaster City Digital Twin: A vision for integrating artificial and human intelligence for disaster management. *International Journal of Information Management* 56:102049.

Fang Z, Gu B, Luo X, Xu Y (2015) Contemporaneous and Delayed Sales Impact of Location-Based Mobile Promotions. *Information Systems Research* 26(3):552–564.

Faraj S, Xiao Y (2006) Coordination in Fast-Response Organizations. *Management Science* 52(8):1155–1169.

Farkas D, Hilton B, Pick J, Ramakrishna H, Sarkar A, Shin N (2016) A tutorial on geographic information systems: A ten-year update. *Communications of the Association for Information Systems* 38(1):9.

Fayyad U, Piatetsky-Shapiro G, Smyth P (1996) The KDD process for extracting useful knowledge from volumes of data. *Commun. ACM* 39(11):27–34, http://dx.doi.org/10.1145/240455.240464.

Felsen P, Lucey P (2017) Body Shots: Analyzing Shooting Styles in the NBA using Body Pose. *MIT Sloan Sports Analytics Conference*.

Fichman RG, Kohli R, Krishnan R (2011) Editorial Overview —The Role of Information Systems in Healthcare: Current Research and Future Trends. *Information Systems Research* 22(3):419–428.

Fonti F, Maoret M (2016) The Direct and Indirect Effects of Core and Peripheral Social Capital on Organizational Performance. *Strat. Mgmt. J.* 37(8):1765–1786.

Fuller A, Fan Z, Day C, Barlow C (2020) Digital Twin: Enabling Technologies, Challenges and Open Research. *IEEE Access* 8:108952–108971.

Ghose A, Li B, Liu S (2019) Mobile Targeting Using Customer Trajectory Patterns. *Management Science* 65(11):5027–5049.

Goebel C, Jacobsen H-A, del Razo V, Doblander C, Rivera J, Ilg J, Flath C, Schmeck H, Weinhardt C, Pathmaperuma D, Appelrath H-J, Sonnenschein M, Lehnhoff S, Kramer O, Staake T, Fleisch E, Neumann D, Strüker J, Erek K, Zarnekow R, Ziekow H, Lässig J (2014) Energy Informatics. *Bus Inf Syst Eng* 6(1):25–31.

Graziano M, Gillingham K (2014) Spatial Patterns of Solar Photovoltaic System Adoption: the Influence of Neighbors and the Built Environment. *Journal of Economic Geography* 15(4):815–839.

Griffiths D, Boehm J (2019) A Review on Deep Learning Techniques for 3D Sensed Data Classification. *Remote Sensing* 11(12):1499.

Gudmundsson J, Horton M (2017) Spatio-Temporal Analysis of Team Sports. *ACM Comput. Surv.* 50(2):1–34.

Gust G, Flath C, Brandt T, Ströhle P, Neumann D (2016) Bringing analytics into practice: evidence from the power sector *International Conference on Information Systems (ICIS)* (Dublin, Ireland).

He D, Ai B, Guan K, Wang L, Zhong Z, Kurner T (2019) The Design and Applications of High-Performance Ray-Tracing Simulation Platform for 5G and Beyond Wireless Communications: A Tutorial. *IEEE Commun. Surv. Tutorials* 21(1):10–27.

Helbich M, Jochem A, Mücke W, Höfle B (2013) Boosting the Predictive Accuracy of Urban Hedonic House Price Models Through Airborne Laser Scanning. *Computers, Environment and Urban Systems* 39:81–92.

Hinz O, Hill S, Kim J-Y, others (2016) TV's Dirty Little Secret: The Negative Effect of Popular TV on Online Auction Sales. *MIS Quarterly* 40(3):623–644.

Holsapple C, Lee-Post A, Pakath R (2014) A unified foundation for business analytics. *Decision Support Systems* 64(1):130–141.

Howcroft J, Kofman J, Lemaire ED (2013) Review of Fall Risk Assessment in Geriatric Populations Using Inertial Sensors. *Journal of Neuroengineering and Rehabilitation* 10(1):91.

Hui SK, Bradlow ET, Fader PS (2009) Testing Behavioral Hypotheses Using an Integrated Model of Grocery Store Shopping Path and Purchase Behavior. *J CONSUM RES* 36(3):478–493.

Hui SK, Inman JJ, Huang Y, Suher J (2013) The Effect of In-Store Travel Distance on Unplanned Spending: Applications to Mobile Promotion Strategies. *Journal of Marketing* 77(2):1–16.

Kane GC, Labianca G (2011) IS Avoidance in Health-Care Groups: A Multilevel Investigation. *Information Systems Research* 22(3):504–522.

Leroy J, Rocca F, Mancas M, Gosselin B (2013) Second Screen Interaction: An Approach to Infer TV Watcher's Interest Using 3D Head Pose Estimation. *Proceedings of the 22nd International Conference on World Wide Web* (ACM Press, New York), 465–468.

Loock C-M, Staake T, Thiesse F (2013) Motivating energy-efficient behavior with green IS: an investigation of goal setting and the role of defaults. *MIS Quarterly*:1313–1332.

Lucey P, Oliver D, Carr P, Roth J, Matthews I (2013) Assessing team strategy using spatiotemporal data. Ghani R, Senator TE, Bradley P, Parekh R, He J, Grossman RL, Uthurusamy R, Dhillon IS, Koren Y, eds. *Proceedings of the 19th ACM SIGKDD international conference on Knowledge discovery and data mining* (ACM, New York, NY, USA), 1366–1374.

Manfreda S, McCabe M, Miller P, Lucas R, Pajuelo Madrigal V, Mallinis G, Ben Dor E, Helman D, Estes L, Ciraolo G, Müllerová J, Tauro F, Lima M de, Lima J de, Maltese A, Frances F, Caylor K, Kohv M, Perks M, Ruiz-Pérez G, Su Z, Vico G, Toth B (2018) On the Use of Unmanned Aerial Systems for Environmental Monitoring. *Remote Sensing* 10(4):641.

Martins-Filho C, Bin O (2005) Estimation of Hedonic Price Functions via Additive Nonparametric Regression. *Empirical Economics* 30(1):93–114.

Menardi G, Torelli N (2014) Training and Assessing Classification Rules with Imbalanced data. *Data Mining and Knowledge Discovery* 28(1):92–122.

Mennecke BE, Crossland MD, Killingsworth BL (2000) Is a map more than a picture? The role of SDSS technology, subject characteristics, and problem complexity on map reading and problem solving. *MIS Quarterly* 24(4):601–629.

Molchanov P, Yang X, Gupta S, Kim K, Tyree S, Kautz J (2016) Online Detection and Classification of Dynamic Hand Gestures with Recurrent 3D Convolutional Neural Networks. *2016 IEEE Conference on Computer Vision and Pattern Recognition (CVPR)* (IEEE, ), 4207–4215.

Perez G, Arbelaez P (2020) Automated lung cancer diagnosis using three-dimensional convolutional neural networks. *Med Biol Eng Comput* 58(8):1803–1815, https://link.springer.com/article/10.1007/s11517-020-02197-7.

Rausch B, Mayer K, Arlt M-L, Gust G, Staudt P, Weinhardt C, Neumann D, Rajagopal R (2020) An Enriched Automated PV Registry: Combining Image Recognition and 3D Building Data. *NeurIPS 2020 Workshop Tackling Climate Change with Machine Learning*, https://arxiv.org/pdf/2012.03690.

Reagans R (2011) Close Encounters: Analyzing How Social Similarity and Propinquity Contribute to Strong Network Connections. *Organization Science* 22(4):835–849.

Rook DW (1987) The Buying Impulse. *J CONSUM RES* 14(2):189.

Rosemann M, Becker J, Chasin F (2021) City 5.0. *Bus Inf Syst Eng* 63(1):71–77.

Saitz, RL, Cervenka, EJ (2019) *Handbook of Gestures: Colombia and the United States* (Walter de Gruyter GmbH & Co KG, Berlin).

Sarkar A (2007) *GIS applications in logistics: A literature review*.

Sevilla J, Townsend C (2016) The Space-to-Product Ratio Effect: How Interstitial Space Influences Product Aesthetic Appeal, Store Perceptions, and Product Preference. *Journal of Marketing Research* 53(5):665–681.

Sharma R, Mithas S, Kankanhalli A (2014) Transforming decision-making processes: a research agenda for understanding the impact of business analytics on organisations. *Eur J Inf Syst* 23(4):433–441, https://link.springer.com/article/10.1057/ejis.2014.17.

Shearer C (2000) The CRISP-DM Model: The New Blueprint for Data Mining. *Journal of data warehousing* 5(4):13–22.

Shirai, Y (2012) *Three-dimensional computer vision* (Springer Science & Business Media).

Shmueli G, Koppius OR (2011) Predictive Analytics in Information Systems Research. *MIS Quarterly* 35(3):553–572.

Stoter J, Kluijver H de, Kurakula V (2008) 3D Noise Mapping in Urban Areas. *International Journal of Geographical Information Science* 22(8):907–924.

van der Pol D, Cuijpers RH, Juola JF (2011) Head Pose Estimation for a Domestic Robot. Billard A, Kahn P, Adams JA, Trafton G, eds. *Proceedings of the 6th International Conference on Human-robot Interaction - HRI '11* (ACM Press, New York, New York, USA), 277.

Venkatesh V, Thong JYL, Xu X (2012) Consumer Acceptance and Use of Information Technology: Extending the Unified Theory of Acceptance and Use of Technology. *MIS Quarterly* 36(1):157–178.

Vrechopoulos AP, O'Keefe RM, Doukidis GI, Siomkos GJ (2004) Virtual store layout: an experimental comparison in the context of grocery retail. *Journal of Retailing* 80(1):13–22.

Wang Q, Kim M-K (2019) Applications of 3D point cloud data in the construction industry: A fifteen-year review from 2004 to 2018. *Advanced Engineering Informatics* 39:306–319.

Wang Z, Li H, Rajagopal R (2020) Urban2Vec: Incorporating Street View Imagery and POIs for Multi-Modal Urban Neighborhood Embedding. *AAAI* 34(01):1013–1020.

Watson, Boudreau, Chen (2010) Information Systems and Environmentally Sustainable Development: Energy Informatics and New Directions for the IS Community. *MIS Quarterly* 34(1):23.

Williams D, Contractor N, Poole MS, Srivastava J, Cai D (2011) The Virtual Worlds Exploratorium: Using Large-Scale Data and Computational Techniques for Communication Research. *Communication Methods and Measures* 5(2):163–180.

Yang D, Kleissl J, Gueymard CA, Pedro HT, Coimbra CF (2018) History and trends in solar irradiance and PV power forecasting: A preliminary assessment and review using text mining. *Solar Energy* 168(Part 1):60–101.

Yu S-M, Han S-S, Chai C-H (2007) Modeling the Value of View in Real Estate Valuation: A 3-D GIS Approach. *Environment and Planning B: Urban Analytics and City Science*:139–153.

Zhang P, Li N, Scialdone M, Carey J (2009) The intellectual advancement of human-computer interaction research: A critical assessment of the MIS literature (1990-2008). *THCI* 1(3):55–107.

# 3D Analytics: Opportunities and Guidelines for Information Systems Research
# Online appendix

## A. Technologies for Capturing 3D Data

Table 9 summarizes the main technologies for capturing 3D data. Among these data generating technologies, light-detection and ranging (LiDAR) and stereo-photogrammetry are typically used to create 3D data sets covering large study areas, such as cities or regions. Since these technologies are able to capture data on large-scale infrastructures, such as buildings or electricity networks, they can provide the basis for IS applications in the areas of smart cities and Green IS. Time-of-flight (TOF) cameras and 3-axis accelerometers are suitable for data collection on smaller scales. TOF cameras provide 3D images of local settings while 3-axis accelerometers detect 3D motions, typically integrated into wearables. A prime application area of these technologies is human-computer interaction, but they also offer opportunities in healthcare IS and sports analytics.

*Table 9 Technologies for capturing 3D data*

| Technology | Data Capturing Process | Application areas | Pros and Cons |
|---|---|---|---|
| **Light detection and ranging (LiDAR)** | LiDAR systems emit laser beams to objects. Sensors capture the reflected signals and elapsed time and derives distance measurements. | High-resolution map creation in remote sensing | Broad range of applications; however, trade-off between range and precision |
| | | Generation of 3D city models with airborne LiDAR. | Small-scale systems (e.g. in handhelds) are low cost |
| | | Navigation for autonomous driving and robotics | Can be used to "see through" certain objects (tree canopy, glass, water) |
| | | Augmented reality applications for tablets/phones | Snow, rain, fog may interfere with measurements |
| **Stereo-photogrammetry** | Objects are photographed from multiple angles with a large overlap (70-80%). The photos can then be triangulated to create 3D representations. | 3D city models (e.g. Google Earth, Google Maps) | Use of standard cameras is possible, therefore less costly than using LiDAR and TOF cameras |
| | | | Requires a large number of overlapping photographs for a single object |
| | | | Matching of photographs to create 3D objects is computationally expensive |
| | | | If multiple 3D objects are in a scene, they can block each other, resulting in missing or incorrect data, requires frequently manual clean up or interpolation |
| **Time-of-flight (TOF) camera** | TOF cameras illuminate a scene and, for each pixel, measure the time of flight of the light to the target object and back to the sensor. | Healthcare (e.g. patient-position and patient-movement monitoring) | In contrast to LiDAR, an entire scene can be measured at once. This enables high refresh rates (hundreds of images per second) and real-time applications. |
| | | Gaming (Kinect) | |
| | | Human-computer interaction | Low-cost (tens-hundreds of USD) |
| | | Robotics | Range is limited to a maximum of 40 m |
| 3-axis accelerometer | Sensor detects motion in the 3D space based on gravity and induced acceleration | Wearables (smartwatches, fitness bands), smartphones, gaming controllers (e.g. Nintendo Wii) | Very small |
| | | | Commodities (cost less than 10 USD) |
| | | | Requires motion, not applicable to static settings |

### B. Characteristics of 3D Data Formats and Associated Modeling Libraries

**Table 1** compares the common 3D data formats and lists popular modeling libraries. Point clouds are the raw data format generated by most 3D data capturing technologies. 3D point clouds are easy to plot, modify and highly interoperable. Since 3D polygon meshes are a more complex data format, there are much fewer modeling libraries available. However, polygon meshes are more informative and generally require less storage than high-resolution point clouds.

*Table 10 Advantages and drawbacks of 3D formats, as well as common modeling libraries*

| Data Format | Pros | Cons | Modeling libraries (R, Python) |
|---|---|---|---|
| **3D point cloud** | Variable level of detail (can be reduced by eliminating or aggregating points) | Often large data sizes (depending on resolution) | Laspy, lidR (preprocessing) |
| | Most basic data format, interoperable, large number of software and libraries exists | | Vedo, PointCloudViewer (visualization) |
| | | | Blender (3D animations) |
| | Plotting possible with most standard libraries | | PyTorch Geometric, Tensorflow 3D (predictive modeling) |
| **3D polygon meshes** | Potentially more storage efficient than point clouds | More complex data format than point clouds (larger number of different components) | Vedo, PyMesh, Rgl (visualization, preprocessing) |
| | | | Blender (3D animations) |
| | | Fewer modeling libraries | PyTorch Geometric, Tensorflow 3D (predictive modeling) |
| | More informative than point clouds | | |

### C. Descriptive Statistics

*Table 11 Descriptive statistics of features in the showcase on real estate valuation*

| Feature | Mean | Median | Stdev | Min | Max |
|---|---|---|---|---|---|
| *elevation above sea level* [m] | 39.39 | 35.61 | 7.39 | 30.96 | 61.54 |
| *latitude* | | | geo-coordinate | | |
| *longitude* | | | geo-coordinate | | |
| *urban district* | | | dummy variables ∈ {0,1} | | |
| *building volume* [$m^3$] | 90106 | 70 | 43173 | 156611 | 1130929 |
| *apartment size* [$m^2$] | 83 | 74 | 39 | 20 | 482 |
| *number of rooms* | 2.52 | 2.00 | 1.01 | 1.00 | 7.00 |
| *apartment rent* [€ / month] | 1068 | 924 | 599 | 220 | 5000 |

*Table 12 Descriptive statistic of continuous features in the showcase on PV adoption*

| Feature | Mean | Median | Stdev | Min | Max |
|---|---|---|---|---|---|
| PV potential [h/a] | 782.4 | 802.6 | 115.7 | 365.1 | 980.6 |
| latitude | | | | geo-coordinate | |
| longitude | | | | geo-coordinate | |
| roof orientation [°] | 144.8 | 144.9 | 117.8 | 0.0 | 360.0 |
| roof inclination [°] | 27.0 | 26.3 | 17.5 | 0.0 | 84.6 |
| roof surface [m²] | 70.8 | 38.8 | 250.5 | 0.0 | 12 59 |

**5.1**

*Table 13 Descriptive statistics of categorical features in the showcase on PV adoption*

| Feature | Level | % |
|---|---|---|
| roof type | flat | 17.0 |
| | A shaped | 67.8 |
| | other | 15.2 |
| building density | high | 16.6 |
| | medium | 51.6 |
| | low | 31.8 |
| neighborhood | historic | 16.6 |
| | residential | 61.2 |
| | commercial | 4.9 |
| | industrial | 6.0 |
| | mixed | 11.3 |
| municipality | A | 65.4 |
| | B | 34.6 |
| PV system | present | 2.0 |
| | absent | 98.0 |
| building function | main building | 21.6 |
| | outbuilding | 13.0 |
| | housing | 33.2 |
| | business | 26.7 |
| | public | 1.4 |
| | other | 4.1 |

### i. Spatial Error Model

We describe the spatial error model (SEM) as specified in the showcase on real estate valuation. In the PV showcase, we employ the model analogously, except for a logit likelihood function.

Since the range of potential spatial effects is not known, we construct a hierarchical Bayesian SEM that allows us to infer the range parameters $(\sigma, \varphi)$ of the spatial error term from the data. The model is composed of a Gaussian likelihood

$$rent_i = Normal(\mu_i, \varepsilon),$$

whose mean $\mu_i$, is linearly related to the predictor variables and an error term $w(s)$.

$$\mu_i = \alpha + \beta_X X_i + w(s).$$

The error term $w(s)$ follows a Gaussian process $w(s) \sim MVNormal(0, K(s, t, \sigma, \varphi))$ with a spatial covariance matrix $K(\cdot)$. The entries of the matrix consist of covariance functions $C$ that model the spatial correlation between sites $s$ and $t$ as a distance-dependent exponential decay

$$C(s, t, \sigma, \varphi) = \sigma^2 e^{-\varphi \, ||s-t||}.$$

We fit the predictor with a Gaussian predictive process (Banerjee et al. 2008) using the framework of Finley et al. (2015). Otherwise, sampling the parameters $\sigma, \varphi$ of the spatial process $w(s)$ would require repeated inversions of $K$, which is of cubic complexity and becomes prohibitive for datasets in the range of $n \approx 10^3$ and larger.

### ii. Computational Model of Solar Irradiance

The computational model calculates the potential yearly electricity that PV systems could generate at each roof polygon, taking into account roof geometries, shading from the environment, and local climate. The procedure comprises four fundamental parts:

2. The potential overall irradiance reaching the study areas is determined using satellite measurements of historical climate and cloudiness.
3. To obtain the irradiance at the roof surfaces, the overall irradiance needs to be transposed to the inclination and orientation of the individual roof polygons.
4. Potential shading is considered, taking into account the position of the sun, surrounding buildings, and the terrain. For this purpose, we apply a ray tracing technique (Corripio 2021) to identify where obstacles (i.e. other buildings or terrain) in our 3D model reside in the path of the sun vector over the course of the day.
5. The resulting irradiance is translated into the power output that a PV system could achieve if it was installed on the particular roof surface. Since PV systems suffer temperature-dependent efficiency losses (Huld et al. 2010), we also take into account historical temperatures.

A video that illustrates solar irradiance and shadows cast over an excerpt of the study area for a sample day can be downloaded here (works best with the VLC media player):

https://www.mediafire.com/file/tztwexsb1abevhf/solar_irradiance_computation.mp4/file.

i. References for the Identification of Research Gaps

Table 14 lists key review articles and seminal papers that informed the identification of research gaps to be addressed with 3D analytics.

*Table 14 Key literature for the development of the 3D analytics research agenda*

| Marketing and Consumer behavior | Healthcare IS | Green IS and Energy Informatics | Smart Cities | Human-Computer Interaction | Sports Analytics |
|---|---|---|---|---|---|
| Andrews et al. (2016) | Agarwal et al. (2010) | Watson et al. (2010) | Marrone and Hammerle (2018) | Zhang et al. (2009) | Xiao et al. (2017) |
| Fang et al. (2015) | Fichman et al. (2011) | Malhotra et al. (2013) | Brandt et al. (2021) | Li and Zhang (2005) | Gruettner (2019) |
| Ghose et al. (2019) | Bardhan et al. (2020) | Goebel et al. (2014) | Giffinger et al. (2007) | Zhang et al. (2002) | Tan et al. (2017) |
| Fong et al. (2015) | | | | | |

References

Agarwal R, Gao G, DesRoches C, Jha AK (2010) Research Commentary —The Digital Transformation of Healthcare: Current Status and the Road Ahead. *Information Systems Research* 21(4):796–809.

Andrews M, Luo X, Fang Z, Ghose A (2016) Mobile Ad Effectiveness: Hyper-Contextual Targeting with Crowdedness. *Marketing Science* 35(2):218–233.

Banerjee S, Gelfand AE, Finley AO, Sang H (2008) Gaussian Predictive Process Models for Large Spatial Data Sets. *Journal of the Royal Statistical Society: Series B (Statistical Methodology)* 70:825–848.

Bardhan I, Chen H, Karahanna E (2020) Connecting systems, data, and people: A multidisciplinary research roadmap for chronic disease management. *MIS Quarterly* 44(1):185–200.

Brandt T, Wagner S, Neumann D (2021) Prescriptive analytics in public-sector decision-making: A framework and insights from charging infrastructure planning. *European journal of operational research* 291(1):379–393.

Corripio JG (2021) insol: Solar Radiation. *R package*, http://CRAN.R-project.org/package=insol.

Fang Z, Gu B, Luo X, Xu Y (2015) Contemporaneous and Delayed Sales Impact of Location-Based Mobile Promotions. *Information Systems Research* 26(3):552–564.

Fichman RG, Kohli R, Krishnan R (2011) Editorial Overview —The Role of Information Systems in Healthcare: Current Research and Future Trends. *Information Systems Research* 22(3):419–428.

Finley AO, Banerjee S, Gelfand AE (2015) spBayes for Large Univariate and Multivariate Point-Referenced Spatio-Temporal Data Models. *Journal of Statistical Software* 63(13).

Fong N, Fang Z, Luo X (2015) Geo-Conquesting: Competitive Locational Targeting of Mobile Promotions. *Journal of Marketing Research* 52(5):726–735.

Ghose A, Li B, Liu S (2019) Mobile Targeting Using Customer Trajectory Patterns. *Management Science* 65(11):5027–5049.

Giffinger R, Fertner C, Kramar H, Meijers E, others (2007) City-ranking of European medium-sized cities. *Cent. Reg. Sci. Vienna UT*:1–12.

Goebel C, Jacobsen H-A, del Razo V, Doblander C, Rivera J, Ilg J, Flath C, Schmeck H, Weinhardt C, Pathmaperuma D, Appelrath H-J, Sonnenschein M, Lehnhoff S, Kramer O, Staake T, Fleisch E, Neumann D, Strüker J, Erek K, Zarnekow R, Ziekow H, Lässig J (2014) Energy Informatics. *Bus Inf Syst Eng* 6(1):25–31.

Gruettner A (2019) What we know and what we do not know about digital technologies in the sports industry *Americas Conference on Information Systems (AMCIS)* ,

Huld T, Gottschalg R, Beyer HG, Topic M (2010) Mapping the Performance of PV Modules, Effects of Module Type and Data Averaging. *Solar Energy* 84(2):324–338.

Li L, Zhang P (2005) The Intellectual Development of Human-Computer Interaction Research: A Critical Assessment of the MIS Literature (1990-2002). *Journal of the Association for Information Systems* 6(11):227–292.

Malhotra A, Melville NP, Watson RT (2013) Spurring Impactful Research on Information Systems for Environmental Sustainability. *MIS Quarterly* 37(4):1265–1274.

Marrone M, Hammerle M (2018) Smart cities: A review and analysis of stakeholders' literature. *Bus Inf Syst Eng* 60(3):197–213.

Tan FTC, Hedman J, Xiao X (2017) Beyond 'Moneyball' to Analytics Leadership in Sports: An Ecological Analysis of FC Bayern Munich's Digital Transformation *AMCIS 2017 Proceedings* , , 10.

Watson, Boudreau, Chen (2010) Information Systems and Environmentally Sustainable Development: Energy Informatics and New Directions for the IS Community. *MIS Quarterly* 34(1):23.

Xiao X, Hedman J, Tan FTC, Tan C-W, Lim ETK, Clemenson T, Henningsson S, Mukkamala RR, Vatrapu R, van Hillegersberg J (2017) Sports digitalization: An overview and A research agenda. *International Conference On Information (ICIS)* ,

Zhang P, Benbasat I, Carey J, Davis F, Galletta D, Strong D (2002) Human-Computer Interaction Research in the MIS Discipline. *Communications of the Association for Information Systems* 9(20):334–355, https://experts.syr.edu/en/publications/human-computer-interaction-research-in-the-mis-discipline.

Zhang P, Li N, Scialdone M, Carey J (2009) The intellectual advancement of human-computer interaction research: A critical assessment of the MIS literature (1990-2008). *THCI* 1(3):55–107.